\documentclass[a4paper,amsmath,superscriptaddress,twocolumn,prb,aps]{revtex4}
\usepackage{graphicx}
\usepackage{epsfig}
\usepackage{color}

\newcommand{\be}{\begin{equation}}
\newcommand{\ee}{\end{equation}}
\newcommand{\bea}{\begin{eqnarray}}
\newcommand{\eea}{\end{eqnarray}}
\newcommand{\ba}{\begin{eqnarray*}}
\newcommand{\ea}{\end{eqnarray*}}
\begin{document}

\title{Ballistic nanofriction}

\author{Roberto Guerra}
\email{robguerra@unimore.it}
\affiliation{CNR-INFM National Research Center S3 and Department
of Physics, \\University of Modena and Reggio Emilia, Via Campi
213/A, 41100 Modena, Italy}
\author{Ugo Tartaglino}
\affiliation{Scuola Internazionale Superiore di Studi Avanzati
(SISSA) and \\CNR-IOM Democritos National Simulation Center, via Beirut 2-4, 34014 Trieste, Italy}
\author{Andrea Vanossi}
\email{vanossi@sissa.it}
\affiliation{Scuola Internazionale Superiore di Studi Avanzati
(SISSA) and \\CNR-IOM Democritos National Simulation Center, via Beirut 2-4, 34014 Trieste, Italy}
\affiliation{CNR-INFM National Research Center S3 and Department
of Physics, \\University of Modena and Reggio Emilia, Via Campi
213/A, 41100 Modena, Italy}
\author{Erio Tosatti}
\email{tosatti@sissa.it}
\affiliation{Scuola Internazionale Superiore di Studi Avanzati
(SISSA) and \\CNR-IOM Democritos National Simulation Center, via Beirut 2-4, 34014 Trieste, Italy}
\affiliation{International Center for Theoretical Physics (ICTP),
Strada costiera 11, 34104 Trieste, Italy}

\date{\today}

\maketitle

{\bf Sliding parts in nanosystems such as Nano ElectroMechanical Systems (NEMS) and nanomotors\cite{Gimzewski,
Urbakh_Nature, vanDelden, Feringa, kim, Balzani, Fleishman, ternes, VanossiPRE}, increasingly involve large speeds, and rotations as well as translations of the moving surfaces; yet, the physics of high speed nanoscale friction is so far unexplored.
Here, by simulating the motion of drifting and of kicked Au clusters on graphite -- a workhorse system of experimental relevance\cite{bardotti2, luedtke, lewis, maruyama}--
we demonstrate and characterize a novel ``ballistic'' friction regime at high speed, separate from drift at low speed. The temperature dependence of the cluster slip distance and time, measuring friction, is opposite in these two regimes, consistent with theory. Crucial to both regimes is the interplay of rotations and translations, shown to be correlated in slow drift but
anticorrelated in fast sliding. Despite these differences, we find the velocity dependence of ballistic friction to be, like drift, viscous.}

\begin{figure}[!b]
\epsfig{file=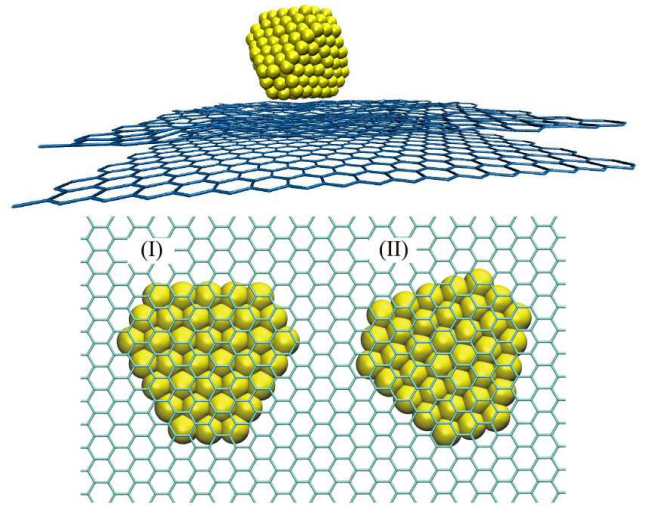,width=8.5cm,angle=0}
\caption{\label{model} {\bf Simulated gold cluster on graphite substrate.} Top panel: A truncated octahedron Au$_{459}$ cluster sliding with its 36-atom $(111)$ facet (area ~ 177 \AA$^2$) over a mobile graphite substrate. Bottom panel: In-registry (I) and out-of-registry (II) geometries for the cluster's $(111)$ contact facet and the graphite substrate.}
\end{figure}

The friction of adsorbed molecules and mobile clusters thermally diffusing on surfaces is gradually moving from a theoretical concept
to reality as shown for example by recent Helium-3 spin-echo experiments~\cite{hedgeland}. As well known in Brownian
motion~\cite{brownian}, everything that diffuses will slowly drift and exhibit viscous friction under a weak force;
we are however much more ignorant about the frictional behaviour to expect when the adsorbate is forced to move at a higher externally imposed speed. High speed friction has long been known in the traditional macroscopic
context\cite{bowden}, and also in AFM experiments\cite{bhushan2005} on asperity dominated rough surface friction, with highly
case-specific outcomes. Here we address the simpler and more fundamental question of speed dependent friction during sliding motion of nanosized objects on atomically flat surfaces, a situation more likely to yield a result of generic validity. To gain experience, we undertake to explore the problem by simulations. Adsorbed Au clusters on graphite\cite{jensen_rev} are known to be thermally mobile even at room temperature\cite{bardotti2}, and make an ideal test case. Experimentally, the low speed Au cluster friction is in principle accessible in Quartz Crystal Microbalance (QCM) experiments\cite{Krim, Mistura, pisov}, while high speed friction
could be realized in AFM/STM tip based setups, where it is possible to pursue kicking techniques, either mechanical\cite{schirmeisen, paolicelli} or electrical.
Anticipating experiment, we simulate by realistic molecular dynamics (MD), the speed-dependent friction of small Au clusters (typically 250-500 Au atoms, as in Fig.~\ref{model}) on a fully mobile graphite surface (see Methods and Supplementary Information).

\begin{figure}
\epsfig{file=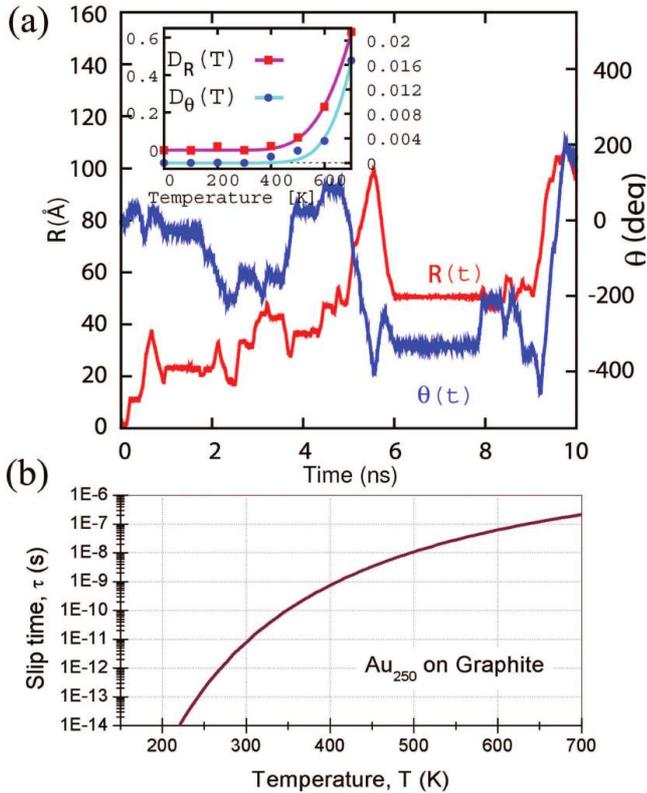,width=8.5cm,angle=0}
\caption{\label{angular_quantization} {\bf Thermal diffusion of deposited Au clusters on graphite.} Panel (a): Force free positional and angular thermal evolution at $T=700$ K.
Pinned states corresponds to parallel orientations of the (111) cluster facet relative to the substrate (see panel (I) in Fig.~\ref{model}).
Note the correlation between positional and angular stops and jumps.
Inset: $T$-dependence of diffusion coefficients (dots), obtained from the first derivative of the mean-square-displacement
for long microcanonical simulations ($>100 $ns). Both positional and angular diffusion fit a thermally activated form $D = D_0\exp\left(-E_0/k_BT\right)$ (continuous lines).
Panel (b):  Predicted temperature dependent QCM diffusional slip time of a drifting Au$_{250}$ cluster on graphite, as expected
by Einstein's relation $\tau =  m D_R/k_B T$  with $D(T)=D_0 \exp(-E_d/k_B T)$,   (experimental
diffusivity values used\cite{bardotti2}, $D_0=10^3$ cm$^2$s$^{-1}$, $E_d=0.5$ eV.}
\end{figure}

Canonical MD is initially used to simulate the force-free cluster diffusion as a function of temperature $T$.
The Au cluster center-of-mass (CM) coordinate $\overrightarrow{R(t)}$ is found to diffuse positionally on graphite, with results
that are close both to experiment and to previous simulations \cite{bardotti2, luedtke, lewis, maruyama}.
In addition we observe that the two dimensional CM positional random walk is closely linked to another conjugate,
and so far unexplored, one dimensional random walk executed by the cluster orientation angle $\theta(t)$ (Fig.~\ref{angular_quantization}(a)).
For relatively long time lapses there is good angular registry between the cluster's $(111)$ contact facet and the
graphite substrate (see bottom panel (I) in Fig.~\ref{model}), and both translations and rotations are stuck--
only small oscillations occur.
Once a thermal fluctuation depins the registry angle $\theta(t)$, the cluster rotates out of commensurability with the substrate, (bottom panel (II) of Fig.~\ref{model}). Now free to slide, the rotated cluster glides long-distance over the surface, with large jumps which dominate diffusion (see Supplementary Information), as recognized long ago by Luedtke {\it et al.} and by others \cite{luedtke, lewis, maruyama}. We find here that both positional and {\it angular} random walks are well fit by $R^2 \sim 4D_{R}t$ and
$\theta^2 \sim 2D_{\theta}t$. For variable temperature, both diffusion coefficients are thermally activated (see inset in Fig.~\ref{angular_quantization}(a)),
$D = D_0\exp\left(-E_0/k_BT\right)$ with $D_{0R} = 0.02$ cm$^{2}$/s, $D_{0\theta} = 0.77\times 10^{14}$
rad$^{2}$/s with activation barriers $E_{0R} = 0.351$ eV and $E_{0\theta} = 0.509$ eV respectively,
underlining the controlling role of angular depinning over positional diffusion\cite{luedtke}. In this respect,
it would be interesting if like positional diffusion in He$_3$ spin-echo \cite{hedgeland}, the angular diffusion of molecules too could be accessed by some experimental technique.


Under a weak applied sliding force, such as realized, e.g., in QCM experiments\cite{Krim, Mistura}, the
freely diffusing clusters or molecules will respond by drifting. Initial simulations in presence of a weak
applied force (see Supplementary Information), show that during drift the strict correlation seen in diffusion between angular depinning and positional jumps is preserved. In QCM one measures the slip time $\tau =  m\mu$, with typical detectable values in the order of $10^{-9}$--$10^{-8}$ sec.
Full fledge drift simulations would of course require prohibitively long times; luckily here we expect drift and diffusion to obey linear response, connecting the positional drift mobility $\mu$ to the viscous diffusion coefficient $D_R$ through Einstein's relation $D_R = \mu k_B T$.  Together, linear response and activated cluster diffusion predict a frictional slip time $\tau = \frac{mD_{0R}}{k_BT} \exp\left(-\frac{E_{0R}}{k_B T}\right)$. Using that,
data of Bardotti {\it et al.}\cite{bardotti2} with a room temperature diffusion coefficient $D \sim 10^{-5}$ cm$^2$/sec for Au$_{250}$ nanoclusters on graphite, translate to a slip time $\tau \sim 10^{-11}$ sec, probably too short for experimental detection, but rapidly growing to measurable values $\sim 10^{-9}$ or larger at $400$ K and above (Fig.~\ref{angular_quantization}(b))~\cite{pisov}.
Actual QCM experiments would be desirable to check this diffusional friction of Au clusters on graphite.


We come now to high speed 
nanofriction, our main task. Here,
the linear response theory connection to diffusion
no longer holds and simulations
are essential. A cluster receiving a large side kick will start off with a large speed $v_0$ and slide 
inertially,
its center of mass slowing down with some as yet unknown time dependence velocity $v(t)$.
We will refer, for lack of a better name, to this frictional regime as ``ballistic'', in analogy with the damped motion of a projectile.
At high speed, the static substrate corrugation is averaged out, and ballistic friction will arise from collisions of the
fast sliding cluster against 
the dynamic excitations of the medium.
The defect free graphite surface has instantaneous and random thermal corrugations, whose magnitude $<u_z^2> \propto k_B T$ at high temperature.
The number of collisions with  these thermal excitations experienced by the fast cluster per unit time should be proportional to speed; and so therefore should the energy damping in linear response.
Supposing therefore provisionally a viscous slowdown (same as for diffusive drift) 
$dv/dt = - \gamma v + \eta$, (here $\gamma$ is a friction coefficient and $\eta$ a random force), friction would imply the exponential decay of cluster speed with time. Thermal corrugations should provide a frictional damping $\gamma_b=F_{friction}/v \propto k_B T$, and thus a "ballistic slip time" $\tau_b= \frac{m}{\gamma_b}$ and a ballistic slip distance $\lambda_b$, (the latter defined as the distance reached by the kicked cluster before major slowdown, equal to $\lambda_b = v\tau_b(1-e^{-1})) \propto \frac{1}{k_B T}$. We therefore expect that the ballistic slip time and distance will drop with temperature, contrary to diffusive sliding, where instead they grew (Fig.~\ref{angular_quantization}(b)). We note however that the significance
of this slip time is quite different from the diffusive one.

We can directly compare this speculative scenario with
actual simulations of high speed cluster sliding. The MD ballistic simulations were initialized by first thermalizing the cluster-surface system at zero speed
and temperature T (for about $100$ ps), then removing temperature control while giving the cluster a large initial sliding speed
$v_0$ (with a typical kickoff value $v_0=100 m/s$, see Methods and movies in Supplementary Information).

\begin{figure}[!b]
 \epsfig{file=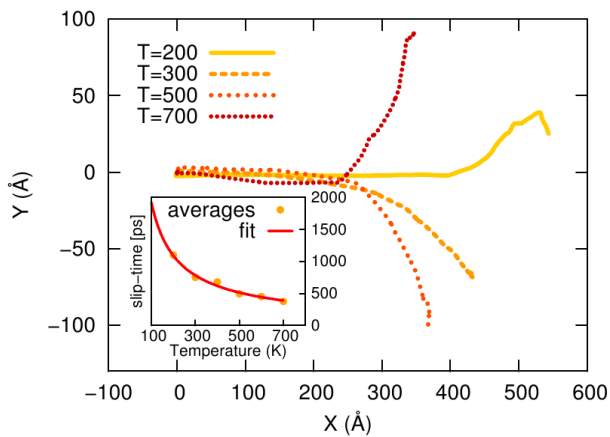,width=8.0cm,angle=0}
 \caption{\label{traiettorie_kicked}{\bf (X,Y) ballistic trajectories of a kicked cluster at different temperatures.} Top view of center of mass trajectories of a Au$_{459}$ cluster
with initial velocity $v=100$ m/s at different temperatures. Note the initial straight ballistic tract, followed by
a broadly curved crossover region. The inset shows the ballistic slip time $\tau_b$, obtained from an exponential fit of the length of the straight
slowdown trajectory portion. The decrease of $\tau_b$ with temperature ($\propto T^{-0.82}$) is close to the expected $1/T$.}
\end{figure}

\begin{figure}[!t]
 \epsfig{file=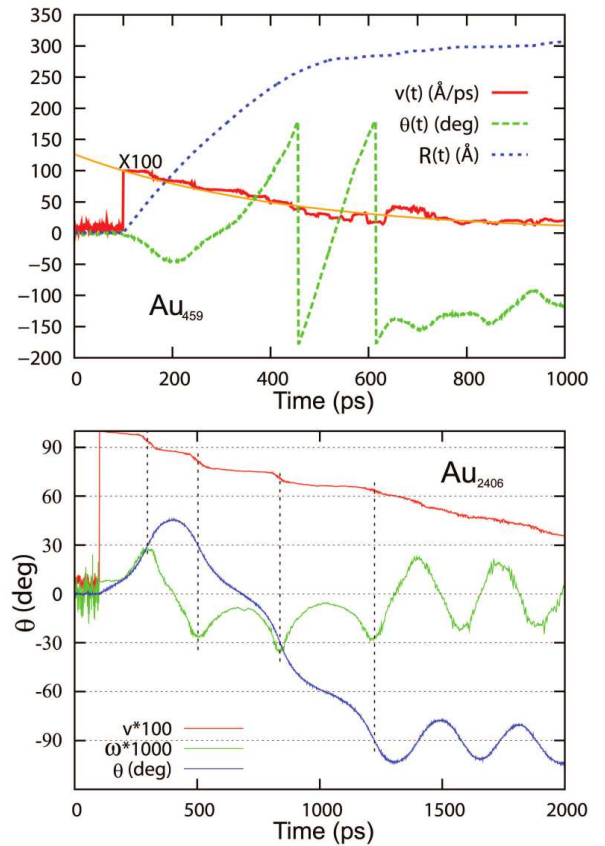,width=8.0cm,angle=0}
 \caption{\label{slip_rot} {\bf Slowdown dynamics of kicked Au clusters.} Top: Slowdown and crossover between ballistic to diffusive friction of a small cluster. CM speed, displacement, and cluster rotational angle are shown. An exponential
velocity slowdown is superposed to the velocity data. At $t=100 ps$, the thermalized cluster is kicked with a high speed $v_0=100 m/s$.  Bottom: detail of ballistic slowdown of a larger Au$_{2406}$ cluster, highlighting clearer sharp drops at exact matching angles of the cluster contact (111) facet with the graphite substrate.
}
\end{figure}

Figure \ref{traiettorie_kicked} portrays the $(X,Y)$ cluster CM trajectories of kicked Au$_{459}$ for increasing temperatures.
The cluster motion is initially ballistic and straight with a slowdown reasonably fit (despite a sizable error
caused by irregular motion) by the viscous exponential law (Fig. \ref{slip_rot}, top panel). The straight trajectory tract
in Fig. \ref{traiettorie_kicked} provides a visual measure of the slip distance $\lambda_b$, with a corresponding slip
time which drops with $T$, close to the expected $1/T$ behaviour (inset of Fig. \ref{traiettorie_kicked}).
The straight ballistic tract generally ends with a sharp turn, followed by a crossover to a curved,
slower and increasingly erratic trajectory.

\begin{figure}[!b]
\epsfig{file=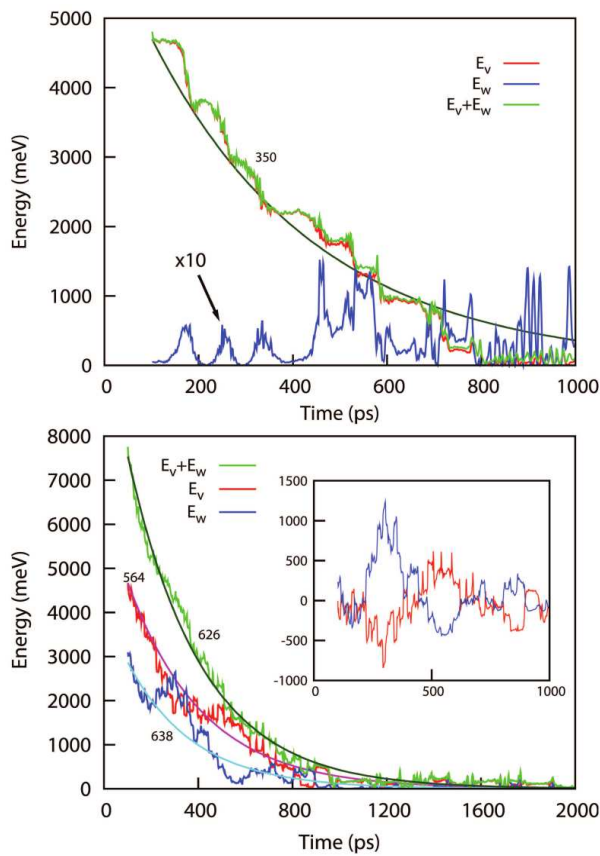,width=8.0cm,angle=0}
\caption{\label{kin_energies} {\bf Time decay of kinetic energies of purely sliding and of sliding/rotating ballistic Au clusters.} Ballistic time dependence of translational (red), rotational (blue), and total (green) kinetic energies of the Au$_{459}$ cluster at $300$ K, initialized with a linear CM velocity $v_0=100 m/s (1$~{\AA}/ps)
and zero initial angular momentum [top], or with an additional large initial rotational speed $\omega_0=0.1$ $rad/ps$ [bottom].
Note the positional-orientational anticorrelation, as CM slowdowns trigger angular speedups. In the rotating case, time decays
of linear, angular and total kinetic energies are fit by exponentials. The inset highlights
the anticorrelation between cluster rotation and translation, obtained by subtracting the smooth exponentials
from $v(t)$, and $\omega(t)$.}
\end{figure}

The role of rotations is important in ballistic frictional slowdown; in fact cluster rotations are closely connected with translations,
as was the case in diffusion and drift. When no angular momentum is initially imparted, the cluster angular dynamics is relatively modest in the straight initial part of trajectories. However, as soon as the cluster-substrate orientations match, the CM translational energy exhibits a sudden drop in favor of an increase of the cluster angular speed, subsequently reversing back to translational, and so on. To make this ``push-pull'' anticorrelation between translations and rotations more visible we simulated clusters with larger moments of inertia and contact facets. MD data for Au$_{2406}$ in bottom panel of Fig. \ref{slip_rot} show a CM slowdown consisting of tilted plateaus for out-of-registry cluster-graphite configurations and sharp drops at accidentally perfect matching angles.
Connections between orientational mismatch and easy sliding was underscored in previous force free diffusion simulations\cite{luedtke}.
In ballistic slowdown the speed plateaus degrade gently, corresponding to ``superlubric'' incommensurate friction,
well known, e.g., for graphite flakes on graphite\cite{dienwiebel}; here, this superlubric dynamical state is favored by the fast sliding motion, preventing significant angular pinning.
As the trajectories cross over from straight to curved, the cluster angular and positional evolutions become
concerted with some analogy with graphite flake behaviour\cite{filippov}. The turning point from a straight to
curved trajectory marks the beginning of a broad crossover between initial ballistic motion and final diffusive
drift (see Supplementary Information).
At the end of the crossover region (whose extension depends on inertia and on T) the, now correlated, stop-and-go motion
typical of diffusion makes its appearance, the cluster-substrate angular registry again corresponding to stops.
This strong translation-rotation interplay is reminiscent of the macroscopic theory on sliding disks \cite{farkas,weidman}, where the friction force and torque are inherently coupled, although differently from here.
%
%
To shed further light on these aspects, we compare the dynamics of clusters initialized with a simple linear velocity $v_0$ to that of
clusters initialized simultaneously with a linear speed $v_0$ {\it and} an angular velocity $\omega_0$ (see movie in Supplementary Information). Their kinetic energy
evolutions are respectively shown in top and bottom panels of Fig. \ref{kin_energies}. On average, both $v(t)$ and $\omega(t)$ decay
in an approximately exponential manner. The plateau-drop-plateau CM motion pattern here disappears, as rotation prevents the touching
surfaces from locking together.
Fluctuations during the ballistic slowdown show a clear {\it anticorrelation} between angular and translational speeds, as highlighted in the figure inset.
This is just the opposite of the correlation found in diffusion and in drift motion, where angular and positional stops and goes coincide.
By analysing the separate translational and rotational kinetic energies we find that in fast rotating clusters the
anticorrelation between translational and rotational speed drops is directly connected with overall kinetic energy near
conservation on a short time scale, resulting in an energy push-pull between the two degrees of freedom. As the skidding
cluster collides with a surface thermal bump it suddenly slows down. The bump is typically not central, and the hit
causes a rotation speedup, with a temporary push of kinetic energy from translation to rotation. The higher rotation
speed now increases the chance to symmetrically hit another bump, whereby kinetic energy is pulled back to translation; and so on.
The total translation plus rotation kinetic energy decreases as a result, as the bottom panel of Fig.\ref{kin_energies} shows,
much more smoothly and exponentially than either of the two components, much bumpier. Ballistic translation-rotation
anticorrelation is also evident in the purely translational data of Fig.\ref{slip_rot} or Fig.\ref{kin_energies}, top panel.
In general, we note that the total kinetic energy decay is well fit by a simple exponential, implying that the overall energy
damping is indeed reasonably described by  $dE/dt \propto -E$, confirming that the ballistic friction of high speed translating-rotating
clusters is again viscous, as in low speed diffusive friction. The reason however seems quite different: simply, in ballistic
friction the number of thermal asperities encountered by the slider must to be proportional to speed.


In summary, the frictional sliding of diffusive clusters (or large molecules, whose behavior should be similar) illustrates two distinct nanofriction regimes -- a standard diffusional one at low speed, and a new ballistic one at high speed --
where the effect of temperature on friction is opposite. Angular motion and rotations cannot be ignored and are crucial to both diffusive and ballistic regimes. The interplay between translations and rotations, besides turning ballistic trajectories from straight to curved, controls the crossover between the two sliding regimes.  Translational and rotational stops and goes, strictly correlated
in diffusion and drift, become anticorrelated in fast ballistic sliding. Ballistic friction, although quite different
from diffusive friction and naturally erratic in small clusters, appears still viscous in its speed dependence,
a useful result whose validity for this and other systems was not discounted in principle, and which it would be interesting
to pursue and test experimentally. Some evidence, we note, may already be present in fast AFM friction data\cite{bhushan2005}.
Angular diffusion might conceivably be addressed by labeling a cluster periphery, and the diffusive drift of clusters
should be directly accessible in QCM experiments~\cite{pisov}. Extending the well developed tip-based
studies~\cite{schirmeisen, paolicelli}, the possibility to kick clusters ballistically either mechanically or electrically
through e.g., a charge pulse, seems entirely open to experimental verification. Finally, advances in fundamental understanding
of ballistic nanofriction should impact practical applications where the relative speeds are not small.

METHODS
%
%
Au fcc clusters were generated and their shape optimized with an empirical EAM potential\cite{eam}. Among several, a truncated octahedron Au$_{459}$ cluster was selected for most sliding simulations (Fig.~\ref{model}). Dynamics of a two-layer and a three-layer fully mobile graphite substrate was considered and described with Tersoff intra-layer potential\cite{tersoff}
and Kolmogorov-Crespi RDP0 interlayer interactions\cite{kolmogorov}. Interatomic Au-C interactions were parametrized by a standard Lennard-Jones form, with parameters yielding realistic corrugation barriers of the graphite substrate against cluster sliding \cite{lewis} (see Supplementary Information).

Temperature equilibration in  force free MD simulations was
enacted by a standard Langevin thermostat. All subsequent frictional simulations were carried out after removing the thermostat, and carefully checking that the frictional energy released into the substrate was small enough to cause a negligible substrate temperature increase. Unlike previous force free simulations\cite{luedtke}, where a rigid substrate sufficed to study diffusion, the fully mobile graphite substrate employed here is essential to study sliding friction. We checked that two or three mobile graphite layers gave sufficiently similar results to mimic the infinite substrate.

Finally, in kicked simulations the initial cluster velocity was given a large value of 100 m/s where the physics of the ballistic regime
is extractable for a substantially lower computational cost than for lower speeds. Results show that crossover from ballistic to diffusive occur between 10 and 1 m/s.

{\bf Acknowledgements}

Discussions with U. Landman, G.E. Santoro, N. Manini, and M. Urbakh are gratefully acknowledged. This work is part of Eurocores Projects
FANAS/AFRI sponsored by the Italian Research Council (CNR), and
FANAS/ACOF. It is also sponsored in part by The Italian Ministry of
University and Research, through PRIN/COFIN contracts 20087NX9Y7 and 2008y2p573. A.V. acknowledges partial financial support by the Regional (Emilia Romagna) Net-Lab INTERMECH.

{\bf Author contributions}

All authors contributed equally to the work presented in this letter.

{\bf Additional information}

Supplementary information accompanies this paper on
www.nature.com/naturematerials. Reprints and permissions information is available online at
http://npg.nature.com/reprintsandpermissions. Correspondence and requests for materials should be addressed to A.V. and E.T.

\end{document}